# Freezing-induced self-assembly of amphiphilic molecules

P. A. Albouy [a], S. Deville [b], A. Fulkar [a,c], K. Hakouk [b], M. Impéror-Clerc [*a], M. Klotz [*b], Q. Liu [a], M. Marcellini [b] and J. Perez [d]

[a] Laboratoire de Physique des Solides, CNRS, Université Paris-Sud, Université Paris-Saclay, 91 400, Orsay, France.
[b] Laboratoire de Synthèse et Fonctionnalisation des Céramiques, UMR3080, CNRS/Saint-Gobain, 550 Av. Alphonse Jauffret, 84306 Cavaillon Cedex, France.
[c] Indian Institute of Science Education and Research, Pune, India
[d] SWING beamline, Synchrotron Soleil, BP 48, 91 192 Gif-sur-Yvette, France

**Abstract**

The self-assembly of amphiphilic molecules usually takes place in a liquid phase, near room temperature. Here, using small angle X-ray scattering (SAXS) experiments performed in real time, we show that freezing of aqueous solutions of copolymer amphiphilic molecules can induce self-assembly below 0°C.

Self-assembly is the process in which objects spontaneously arrange themselves into ordered phases through weak interactions and has emerged as a central notion to understand the organization of matter at the nanoscale. In particular, self-assembly of amphiphilic molecules plays a central role in soft matter, biology, and numerous technological applications. Self-assembly of lipids, surfactants, copolymers, liquid-crystalline molecules, or colloids can be defined as the equivalent of the crystallization phenomenon for atoms or molecules, but at a larger length scale and for objects interacting through weak interactions and not strong ones (like covalent or ionic ones). The story of self-assembly started in the field of biological membranes when it was discovered that lipids can spontaneously form ordered phases, ranging from a simple 1D stack of lipid bilayers in an aqueous medium (lamellar phases) to much more complex 3D periodic architectures such as cubic phases.[1] Cubic phases are now used, for instance, for crystallization of membrane proteins.[2] The field of liquid crystals is certainly a major scientific area based on self-assembly, with many successful technological applications. For example liquid crystal displays (LCD) are based on the self-assembly of liquid-crystalline molecules into corresponding liquid-crystal phases. Such phases exhibit strongly anisotropic properties, used in the devices, because of the specific symmetries of their self-assembled architectures. Self-assembly has also become an inspiring path for bottom-up approaches[3] to obtain many different nano-structured materials, hybrid materials, mesoporous materials, nanoparticle assemblies and so on.

Self-assembly in an aqueous medium is specifically interesting for many practical and biological implications.[4] Here, we report the self-assembly of a copolymer that can be achieved by freezing its aqueous solution below its freezing point. This process is fast, easy to control, and is a new way to induce self-assembly of an amphiphilic molecule. We believe this approach is independent of the copolymer used so far, and could thus be applied to any other type of amphiphilic molecules or building blocks known to organize via self-assembly in a solvent, provided that the freezing point of the solvent is within an easily accessible temperature range at atmospheric pressure.

Using in situ synchrotron Small Angle X-ray Scattering (SAXS), we followed the self-assembly during freezing of aqueous solutions of a commercial block-copolymer composed of ethylene oxide (EO) and propylene oxide (PO) blocks, named P123 ($EO_{20}PO_{70}EO_{20}$), in the concentration range 5–13 wt% (see Methods section). In these solutions, a phase transition occurs at the critical micellar temperature (CMT), from individual P123 molecules called unimers (below the CMT) to aggregates of P123 molecules called micelles (above the CMT). The value of the CMT decreases with increasing copolymer concentration, but it is always higher than 10 °C for the concentration range investigated here (Fig. S1, ESI†).[5] Prior to freezing, the solution was initially cooled down to 5 °C and it was checked by SAXS that no micelles were present in solution, confirming that all the P123 molecules were solubilized as unimers. The bottom of the sample was then cooled below 0°C so as to induce an upward moving freezing front into the sample. Freezing at the bottom starts after 8 minutes as revealed by the temperature increase due to the latent heat release (point 1 in Fig. 1). The subsequent displacement of the freezing front was monitored by a video camera (Fig. S2, ESI†). The transmission of the incoming X-rays beam (Fig. S3, ESI†) and SAXS patterns (Fig. 1C) were simultaneously time-recorded at 10 mm above the sample bottom. The growth of vertical needle-like ice crystals at this position was detected by the appearance of narrow horizontal strikes on the SAXS patterns close to the beamstop attributed to the reflection of the X-ray beam on the crystallites surface (point 2 in Fig. 1, time = 10 min).

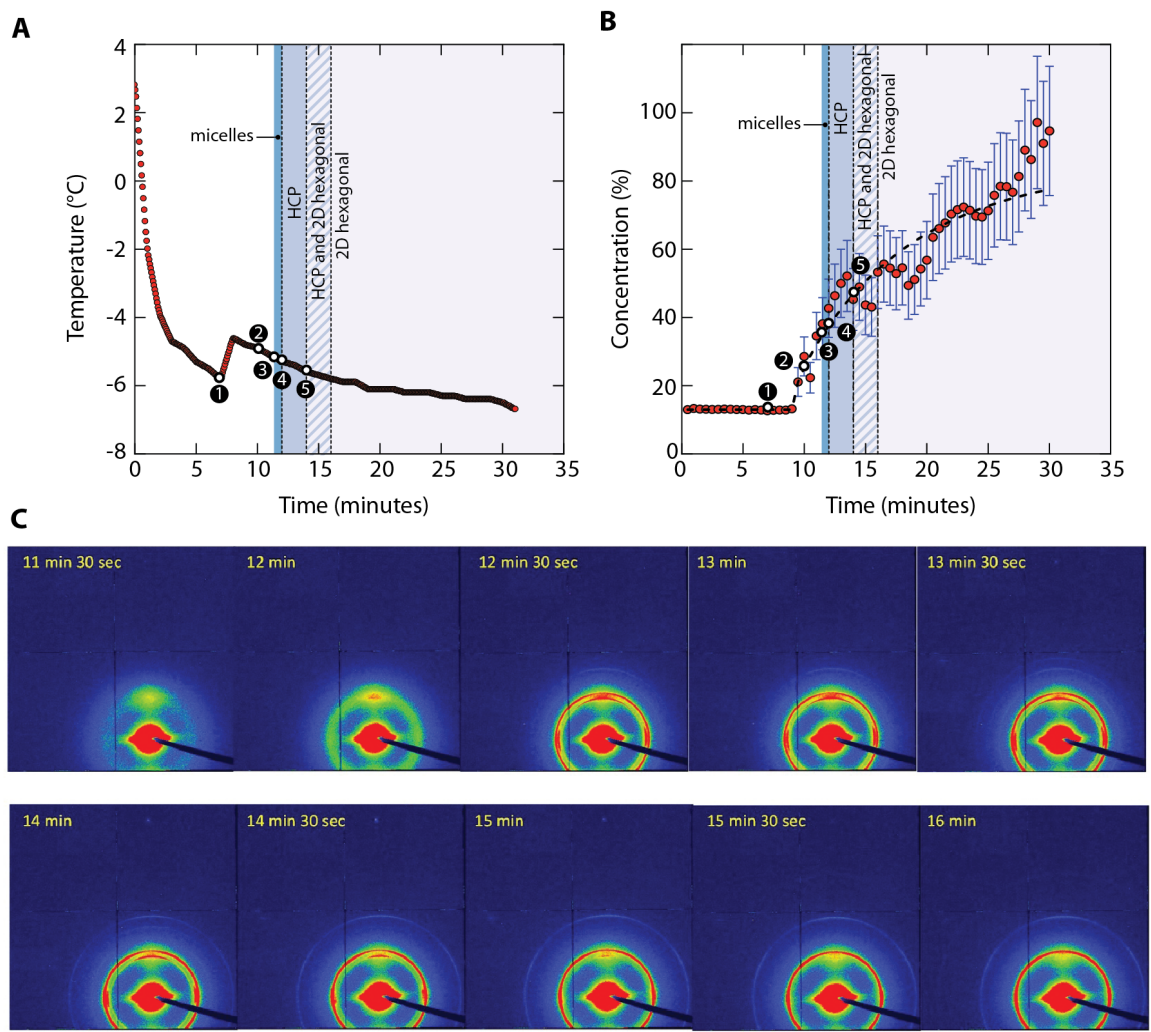

Figure 1: Self-assembly during freezing: evolution of temperature, concentration, and SAXS patterns. Sample concentration of 13 wt%. (A) Evolution of the temperature at the bottom of the sample. The peak at 8 min corresponds to the latent heat release when the first ice crystals are formed. The points 1 to 3 correspond respectively to the nucleation of ice (1), the time when the ice crystals reach the observation window (2) and the appearance of a liquid structure factor that indicates the presence of micelles (3). (B) Increase of the copolymer concentration due to ice formation. (C) Time-lapse evolution of SAXS patterns showing the self-assembly. The following sequence is observed: a concentrated liquid of micelles (structure factor signal) (3), followed by a hcp phase of spherical micelles (4), and a 2D hexagonal phase of cylindrical micelles (5).

The P123 molecules are expelled from the growing ice crystals, effectively increasing the P123 copolymer concentration in the remaining liquid phase. The first pattern in Fig. 1C, at 11.5 minutes, shows two different signals: the horizontal narrow strikes (in red, close to the beamstop) linked to the presence of ice crystals, and a broad ring (green/yellow) more intense along the vertical direction. This ring is a background signal due to the extruded plastic sample holder. The extrusion induces a preferential orientation of the polymer chains in the sample holder that results in an anisotropic SAXS signal. These two signals are therefore present on all patterns afterwards. No specific signal of micelles is present, so we assume that this liquid phase is a mixture of P123 and the part of the water molecules which are still in a liquid state. The actual concentration of the P123 molecules (Fig. 1B) can be derived from the transmission of the sample, by comparison with the transmission of a pure water sample during freezing (Fig. S3, ESI†). The concentration effect of the growing ice-crystals induces a very fast self-assembly process, as revealed by the evolution of the SAXS patterns (Fig. 1C): at 12 minutes, a narrower ring (in green) is detected, and is attributed to the main peak of the structure factor of P123 micellar aggregates, revealing that such aggregates form rapidly within just 30 s.

Several very narrow Bragg peaks are then detected, revealing the ordering of the micelles into three dimensional periodic lattices. The positions of the peaks can be unambiguously attributed to two successive periodic phases (Tables S4 and S5, ESI†). First, a transient phase is observed that consists in a hexagonally close packed (hcp) phase of spherical micellar aggregates (space-group $P6_3/mmc$).[6] Then, a 2D hexagonal packing of cylindrical micelles is stabilized until the end of the freezing experiment (space-group $P6/mmm$). After 30 minutes, the sample was entirely frozen. SAXS was then carried out along the vertical direction (Fig. 2), which corresponds to the growth direction of the ice-front. The 2D hexagonal phase is always present, except at the very top of the sample where the hcp phase is stabilized in a narrow region (Fig. 2C). An important observation is that both phases grow with a preferred orientation. All the domains of the hcp phase have their c-axis perpendicular to the growth direction, as can be deduced from the SAXS pattern (Fig. 2B) which exhibits modulation along the scattering rings. The domains of the 2D-hexagonal phase are preferentially oriented with the a* direction and the cylinder axis perpendicular to the growth direction, as shown by the 6-fold symmetry modulations of the intensity along the 10 scattering ring (Fig. 2A). These preferential orientations could be related to the growth habit displayed by the ice front, which consists of ice needles oriented along the temperature gradient direction (Fig. 2). It is known that the denser planes for a given structure have a tendency to lie parallel to the deposition surface, because the interfacial energy is minimized.[7] In the present case, such planes are the (001) planes for hcp and (10) planes for 2D-hexagonal. Furthermore, the micellar cylinders are lying parallel to the growth front, and it may indicate that a horizontal interface is also inducing orientation during the freezing process.

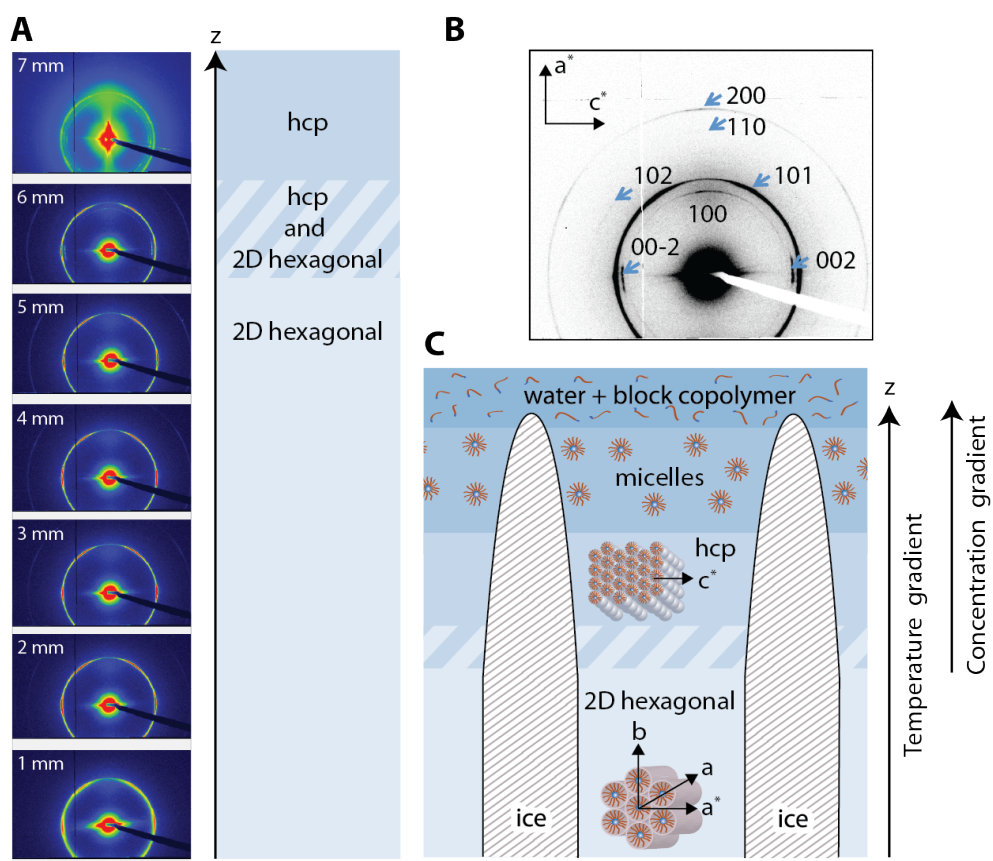

Figure 2: Self-assembly along the direction of the temperature gradient. Sample concentration of 13 wt%. (A) Scan along the direction of the temperature gradient after 30 minutes. The 2D hexagonal phase is present over almost the entire region where ice is present (5 mm in height). At the top, the hcp phase is present only in a limited region of about 1 mm just below the liquid phase region (without ice crystals). (B) SAXS pattern of the hcp phase: several domains of the hcp phase coexist with a preferential orientation of their c axis perpendicular to the growth direction. (C) Sketch of the self-assembly along the growth direction, including the preferential orientation of the domains. The domains of the 2D-hexagonal phase are preferentially oriented with the a* direction perpendicular to the growth direction.

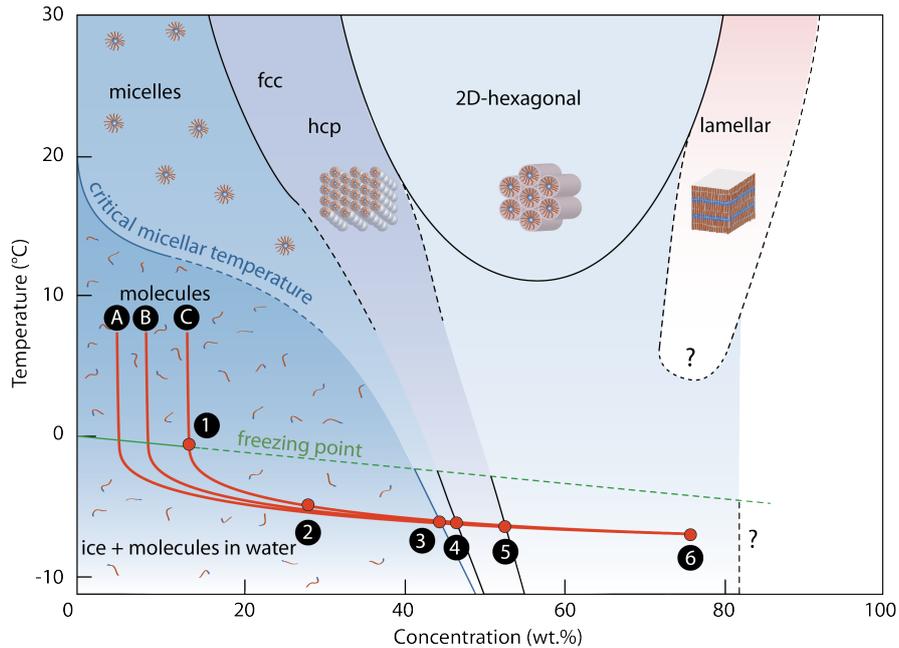

Figure 3: Scheme of the freezing-induced self-assembly of P123 in water in a concentration–temperature diagram. Initially, only P123 unimers are present in solution. As soon as ice-crystals are forming (1-2), the concentration of P123 (inside the liquid phase) is increased and self-assembly takes place (3-4-5). The CMT curve, and the freezing point temperature curve (FP) are extrapolated (dashed lines). Frontiers of the different phases below 15°C are only indicative.

Interestingly, these two types of periodic phases (hcp and 2D-hexagonal) have already been observed in the phase diagram of P123 in water, but in a range near room temperature and well-above 0 °C.[8,9] To our knowledge, a complete phase diagram of P123 in water below room temperature is not available in the literature. The concentration range for which they are observed in such phase diagrams is in good agreement with the observed sequence during freezing, with the hcp phase around 50 wt% and the 2D-hexagonal phase at higher concentrations (Fig. 3). From our experiments, we can conclude that these organized phases can exist at temperatures below the freezing point with no major modifications of their structures. At this point, we cannot conclude whether these phases are frozen, or if liquid water is still present in these concentrated micellar phases. However, the penetration of ice in confined spaces requires much lower temperatures.[10] Liquid water is therefore probably still present in the concentrated micellar phase at the lowest temperatures achieved here (−7°C).

It is generally accepted that self-assembly in aqueous solution of a copolymer, surfactant, lipid or any other object with weak interactions can occur only when micellar aggregates are already present in solution, and consequently cannot take place below the CMT. Our results show that it is not always true, and that micelles can form at temperatures lower than the initial value of the CMT. The formation of micelles is driven by the increase of the copolymer concentration when ice is formed. Two phases coexist in the sample, both containing water molecules: ice crystals and a concentrated aqueous solution of P123. We assume that the water molecules remaining in solution with the P123 molecules are in the liquid-like state, and are linked by H-bonds to the EO groups of the copolymer.[11] The equilibrium between these two phases is fixed by the chemical potential of the water molecules, which is imposed by the ice phase and depends only on temperature. Note that we verified that the freezing-point depression temperature due to the presence of P123 in solution is small, typically of −0.5 °C. The colligative freezing point depression of P123 is therefore smaller than that of other solutes (Methods).[12]

The final P123 concentration depends only on the temperature of the sample and not on the initial P123 concentration. This hypothesis is validated by comparing experiments performed with different initial concentrations and similar final temperatures (Table 1 and Table S5, ESI†). The lattice parameters of the hcp and 2D-hexagonal phase have comparable values for all initial concentrations. This is in good agreement with the previous observations that freezing-induced segregation can increase the concentration of solutes, molecules, or particles by several orders of magnitude.[13,14]

Table 1 Effect of the initial concentration in P123 [P123]$_{ini}$. Estimated concentrations (wt%) of the liquid phase from transmission values (Fig. S3, ESI) are given for different states: micelles, hcp phase and 2D-hexagonal phase. Error bar for concentration is ±5 wt%. The average values of the lattice parameters for the hcp and 2D-hexagonal phase are reported (Table S5, ESI). Error bar for lattice spacing is 0.05 nm

| | | | | | Lattice parameters | |
|---|---|---|---|---|---|---|
| [P123]$_{ini}$ | Micelles | hcp | hcp+2D-hex | 2D-hex (nm) | a (hcp) | a (2D hex) |
| 13 | 43 | 44–46 | 46–52 | 53–80 | 14.86 | 12.94 |
| 10 | 40 | 40–53 | 54 | 54–80 | 14.86 | 12.94 |
| 8 | 50 | 50–63 | 63–65 | 66–80 | 15.23 | 13.2 |
| 5 | 24 | 30–42 | 44 | 47–80 | 14.99 | 13.13 |

Finally, the evolution with time of the system during freezing-induced self-assembly can be summarized using a concentration-temperature diagram (Fig. 3). First, before ice formation, the concentration of the unimers is constant and temperature is decreasing rapidly. As soon as ice forms (freezing point curve), the concentration of the copolymer starts to increase while temperature nearly stabilizes. Micelles form when the concentration crosses the CMT curve (dashed line) and self-assembly of the micelles into ordered mesophases subsequently occurs at still higher concentrations. Indeed the phase diagram of the P123 copolymer in water is not precisely established below room temperature.[8] No information can be found below freezing temperatures, and the frontiers for the phases (fcc, hcp, 2D-hexagonal and lamellar) in this diagram are thus only indicative.

The freezing-induced self-assembly demonstrated here may be important in several domains in soft matter and materials science. A first possibility is the field of ordered mesoporous materials. Indeed, self-assembly has been extensively used for the preparation of mesoporous materials by templating methods. The mechanism evidenced here with freezing recalls the mechanism of evaporation induced self-assembly (EISA) used for the preparation of films,[15] powders,[16] or monoliths,[17] where self-assembly is similarly driven by a concentration effect (Fig. 3). Freezing-induced self-assembly offers a new route for the elaboration of mesoporous materials. Silica[18] and alumina monoliths[19] have already been prepared by the association of ice-templating, P123 self-assembly and sol–gel chemistry to manufacture materials with hierarchical multi-scale porosity. Mesopores were templated by P123 micelles and macropores by ice crystals. The 2D hexagonal structure of the micelles was retained in the final material, providing a 2D hexagonal organization of the mesopores. The observed orientation of the self-assembled micelles with respect to the ice-front should permit a control of the pore orientation for the material, which can be an important issue. This orientation effect due to minimization of surface energy at interfaces is similar to what has been reported for mesoporous materials grown under confinement within anodic alumina membranes.[20]

Freezing provides another pathway for the self-assembly of many other elementary building blocks. The principles shown here with an amphiphilic block copolymer can be extended to other systems where weak interactions can trigger self-assembly when the concentration is high enough. This could be particularly important for systems sensitive to temperature. Unlike evaporation induced self-assembly, which can require temperatures up to 80°C, self-assembly occurs here at low temperature (−5°C). Temperature sensitive building blocks such as lipids could thus be organized with this approach while preserving their integrity. This could be particularly relevant for biological materials. Very stimulating implications could be the formation upon freezing of lipids vesicles linked to prebiotic synthesis theories.[4] Self-assembled vesicles are considered essential components of primitive cells, and prebiotic synthesis from frozen solutions is considered as a serious contender to explain the initial formation of vesicles.[21] The segregation by the ice crystals provides a very efficient mechanism to increase the concentration of the building blocks by several orders of magnitude and avoid an endless dilution of the formed vesicles. Block copolymer vesicles, in particular, have been used for light-driven ATP generation[22] and enzymatic cascade reactions.[23] We expect therefore that our results obtained for freeze-formed micelles can be used in such studies. Investigating whether lipids vesicles of larger dimensions can be formed by freezing is a particularly exciting perspective.

## Methods

Solutions of the P123 copolymer (EO20PO70EO20, BASF®) in water (5, 8, 10 and 13 wt%) were prepared without any purification. SAXS experiments were performed at the SWING beamline of the French SOLEIL synchrotron facility. Transmission and SAXS pattern were recorded during freezing using an incident X-ray beam of wavelength 0.124 nm. Exposure time for SAXS was 2 s, with acquisitions every 30 s. Samples were introduced inside a vertical commercial plastic straw of diameter 7.7 mm filled up to 10 mm. Freezing of the sample was obtained using a commercial device (CT160, Deben company) equipped with a Peltier element. For the measure of the freezing point depression, 100 ml of P123 solution (13 wt%) was inserted inside a Dewar and cooled down until ice crystals were formed. A freezing point depression of −0.5°C was measured when heating the solution.

## Acknowledgements

This project was supported by the European Research Council under the European Community's Seventh Framework Pro-gramme (FP7/2007-2013) grant agreement 278004 (FreeCo). We acknowledge Synchrotron Soleil for allocation of beamtime at the SWING beamline.

## Notes and references

**Footnote**

† Electronic supplementary information (ESI) available: Supplementary figures/Tables S1–S5. See DOI: 10.1039/c6sm02154a